*E.G. Pashinskaya, V.N. Varyukhin, A.A. Maksakova, A.I. Maksakov[1], A.A. Tolpa[1], A.V. Zavdoveev*

# EFFECT OF THE TECHNIQIE OF DRAWING WITH SHEAR ON THE STRUCTURE AND THE PROPERTIES OF LOW-CARBON WIRES

Donetsk Institute for Physics and Engineering named after A.A. Galkin NAS of Ukraine
72 R.Luxemburg Str., 83114, Donetsk, Ukraine

[1]SPA «Donix», Donetsk, Ukraine

To obtain the materials of ultrafine grain (UFG) structure, different methods of severe plastic deformation (SPD) with shear in both cold and hot state are used: equal-channel angular pressing, twist extrusion, and combination of these methods with succeeding rolling, upset, drawing etc. The application of these methods allows substantial increase in the strength of the material at certain conserved reserve of plasticity.

However the mentioned combination of the methods can not be realized at wiredrawing production plants, whereas the last are very interested in new technological and operation characteristics of long-length wire products [1,2]. One of constraints imposed upon the production of UFG wire is that the total output of the materilas produced by the listed SPD methods is measured by tens of kilograms and tons but the required capacity of wiredrawing production is hundreds of thousand tons.

A possible solution of the problem of the long wire products can be application of drawing with shear. As may be supposed, an increase in the plasticity reserve of rods and wires allows cheapening and simplification of the production technique due to the abandonment of the intermediate anneal.

In [1–9], different methods of severe plastic deformation of long metal billets of varied configuration are described. The works [5,6,9] are of special interest here.

The authors of [5] analyze the application of sign-alternation twisting of cold-drawn furniture without an additional heating. The main advantages of the method are continuous operation and possible application to the production of long products with enhanced mechanical properties. In [6], a method of plastic structure formation in the material of long billets is tested, and a facility of the realization is presented that is based on sign-alternating deformation within intersecting channels. The deformation zone is formed within the billet due to a shift of the symmetry axes of the channels combined with uniaxial tension. The presented method is intermittent, and the finite billets of several meters in length can be produced. The advantage of the method is formation of a fine grain structure. But the deforming block is even more complex than that in the case of [5]. Both the methods [5,6] do not permit production of the wire of small diameter.

A drawing-based method of obtaining of UFG structure in long products is described in [9]. The main advantage is continuous operation and possible application to the large-scale drawing production. A drawback is labor intensity of drawing because a complex technical facility is used that requires disassembling and new installation in the course of replacement of wire drawing dies.

The present paper reports the developed technique of drawing with shear aimed at enhancement of technological plasticity of low-carbon steels without thermal treatment. The technology should provide definite physical and mechanical properties of the wire. It also should be cheap, simple and reliable in the course of operation.

**Experimental technique**

The experiment was carried out with using the billet of steel ER70S-6 (Table 1). The drawing was realized at the plant AZTM 7000/1 by the developed (experimental) technique (with shear dies) and by the classical one (with the standard round dies). The drawing routes of the both techniques are listed in Table 2.

Table 1

**Chemical composition of ER70S-6 steel, %**

| C | Mn | Si | S | P | Cr | Ni | Cu | N |
|---|---|---|---|---|---|---|---|---|
| 0.071 | 1.98 | 0.84 | 0.015 | 0.018 | 0.015 | 0.009 | 0.016 | 0.0055 |
| Ti | As | B | Al | V | Mo | W | Co | |
| < 0.005 | < 0.005 | < 0.0005 | 0.005 | 0.006 | < 0.01 | 0.024 | 0.01 | |

Table 2

**Drawing routes for the wires produced by the experimental and the standard techniques**

| Technique | Die diameter, mm | | | | | | | | | |
|---|---|---|---|---|---|---|---|---|---|---|
| Experimental | 6.15 | 5.4 | 5.2* | 5.0 | 4.30 | 3.90 | 3.5 | 3.06 | 2.70 | 2.39 |
| Conventional | | | 5.30 | | | | | | | |

*Note.* * – die with shear

The obtained samples from 6.15 to 3.90 mm in diameter were subjected to mechanical tests, in particular, the ultimate tension strength and the contraction ratio were measured (Fig. 1).

The microstructures of the annealed sample and the deformed ones were studied at the 100–1000 power device «Neophot-32» after repetitive polishing and etching of the grain boundaries (the composition of the etching agent: 4% nitric acid, 97% alcohol). The photos were made by the optical microscope Axiovert 40 MAT.

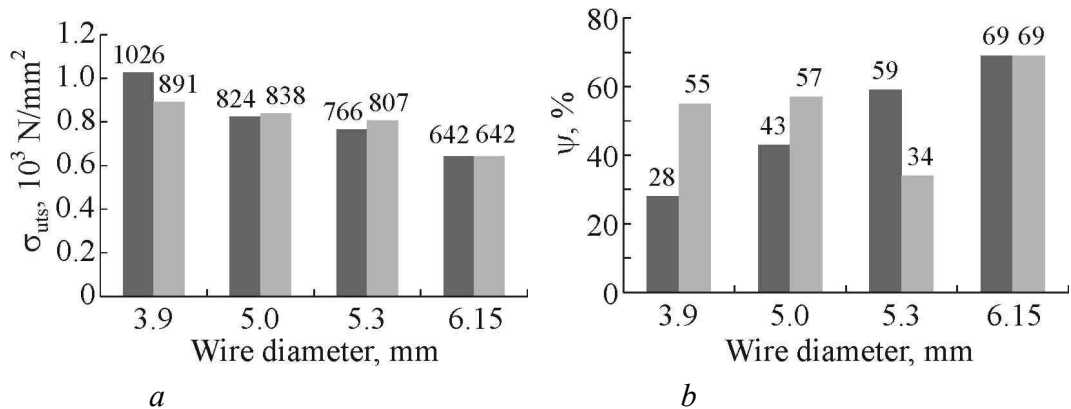

**Fig. 1.** Mechanical properties of the ER70S-6 steel wire obtained by the conventional (■) and the experimental (■) technique: *a* – the ultimate tensile strength $\sigma_{uts}$, *b* – the contraction ratio $\psi$

Table 3

**Elongation ratio *k* of the ER70S-6 wire**

| Technique | k | | |
|---|---|---|---|
| | ⌀6.5 | ⌀5.0 | ⌀3.9 |
| Experimental | 1 | 0.35 | 0.19 |
| Conventional | | 0.21 | 0.10 |

The estimation of the grain size and the fragment size was performed in the transversal and longitudinal directions of the samples. 100 measurements were made on every photo. The elongation ratio was calculated as

$$k = D_1/D_2, \qquad (1)$$

where $D_1$, $D_2$ were the lengths of the grain along the grain elongation and in the cross-section of the sample, respectively, mm.

Besides, Vickers hardness HV (loading of 200 g) and microhardness $H_\mu$ (loading of 100 g) were measured. The measurement error was ±5%. The density of the samples was measured by hydrostatic weighing. Strength tests were carried out with using UMM-50 test machine at the temperature of 293 K and the rate of loading of 10 mm/min according to GOST 25.601–80.

## Experiment results

The tests have demonstrated that when the diameter of the sample is reduced, the ultimate tensile strength remains at the same level irrespective of the conventional or experimental technique (Fig. 1,*a*), and the contraction ratio of the sample processed by the experimental technique is higher as compared to the conventional drawing (Fig. 1,*b*).

On the basis of the mean values, it is seen that the reduction of the diameter after the conventional technique is associated with the increase in the ultimate tensile strength by 380 N/mm$^2$, and the experimental technique yields 240 N/mm$^2$.

After the conventional technique, the value of the contraction ratio is substantially decreased from 69 to 28%, and the drop after the experimental technique is from 69 to 55%.

One of the distinguishing features of the experimental technique is a decrease in the structure anisotropy in the longitudinal samples. An evidence of the fact is an increase in the elongation ratio (Table 3). The photos of microstructures confirm the calculated data (Fig. 2).

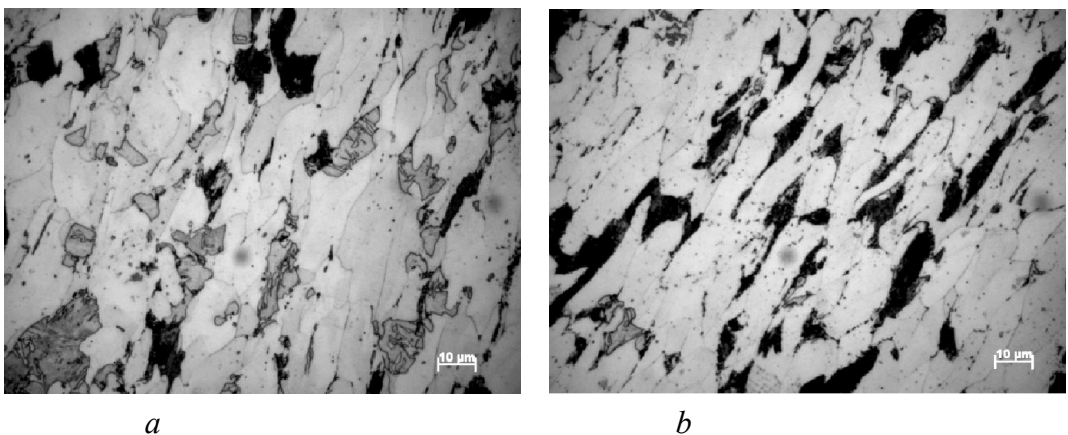

*a*  *b*

**Fig. 2.** Microstructure of the samples of the wire of ER70S-6 ⌀5.0 obtained by the experimental technique (*a*) and the conventional technique (*b*), ×─0. The scale ⊢⊣ represents 10 μm

The analysis of the structures revealed the following principal structure features.

1. The etchability of perlite colonies is higher at the conventional technique that is related to non-equilibrium of perlite after drawing. In the longitudinal section, along the section at the conventional technology is larger that of the experimental technique.

2. It has been established that the experimental technique results in decreased anisotropy of the grains, and the grain size is reduced as a whole, when the degree of deformation increases. But at certain stages, successive decrease and increase in the grain size is observed that is supposed to be related to the progress in competitive processes of fragmentation and dynamical polygonization.

For instance, in the tested samples of ⌀5.2 and ⌀5.0 in cross-section, the structure is enlarged: the ferrite grains and the sizes of perlite colonies grow as compared to the sample of ⌀6.15. In the cross-section of the experimental sample, the structure was larger in comparison with the sample obtained by the conventional technique. The structure of the last sample contained more perlite phase.

3. In the cross-section of the samples of ⌀3.9 obtained by the conventional technique, there is a surface zone of about 200–300 μm that differs in etchability. This effect is not found after the experimental technique, the metal is homogeneous, the ferrite grains are lager, the size and the amount of perlite is much less than that after the conventional technique.

4. The samples of ⌀2.39 by the conventional technique are etched better, so, the degree of non-equilibrium is higher. In the longitudinal section of the conventional samples, the anisotropy is more intensive, the ferrite grains are smaller, the perlite colonies are elongated as stripes (Fig. 3). The same effect is observed in the cross-section. Besides, healing of pores and microcracks is registered in the course of the experimental drawing (compare Fig. 3,*c* and *d*).

## Conclusions

1. The application of the experimental dies allows improvement of the mechanical properties of the samples when the diameter is decreased after drawing: the contraction ratio of the wire is insignificantly reduced and stays at a relatively high level as compared to the conventional technique where the contraction ratio is halved.

2. The use of the experimental technique permits variation of the ferrite grain size (successive increase and decrease) in comparison with the conventional technique where the increase in the degree of deformation results in unambiguous grain reduction.

3. The application of the experimental technique generates pore healing in the wire of a small diameter.

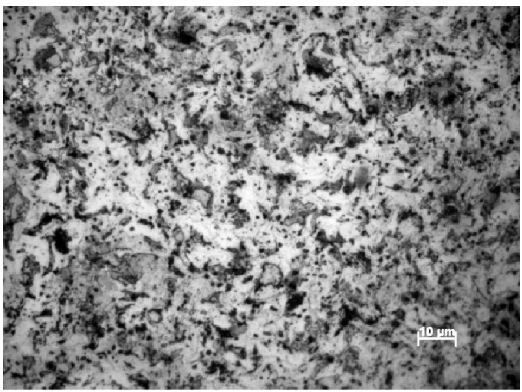 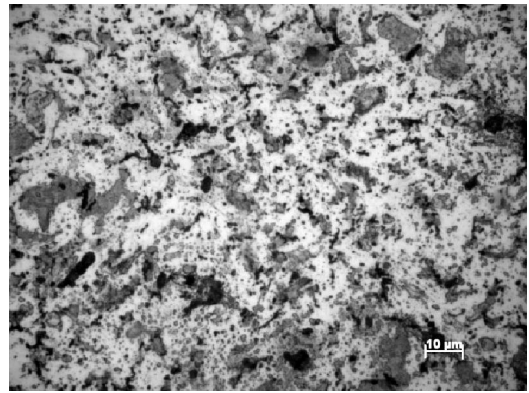

*a*          *b*

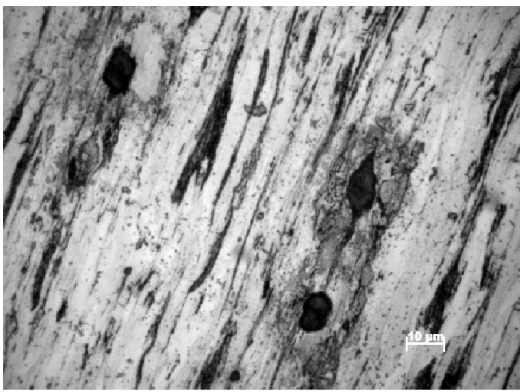 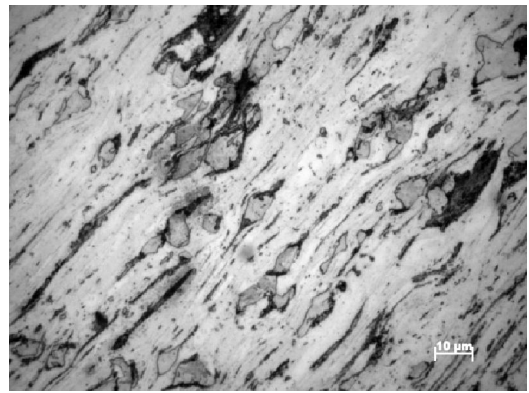

*c*          *d*

**Fig. 3.** Microstructure of the central zone of the ER70S-6 wire samples of ⌀2.39 obtained by th4e conventional technique (*a*, *c*) and by the experimental technique (*b, d*) : *a, b* – the cross-section, *c, d* – the longitudinal section, ×100. The scale is as in Fig. 2

4. To prevent the heating of the wire and the die, it is necessary to pass to the experimental drawing through two experimental dies with shear separated by a standard die. This configuration will result in enhanced fabricability and reduction of the drawing force.